\documentclass[amsmath,amssymb,groupedaddress,superscriptaddress,lengthcheck,aps,prl] {revtex4-1}

\usepackage{graphics}
\usepackage{graphicx}   
\usepackage{bm}
\usepackage[latin1]{inputenc}
\usepackage{textcomp}
\usepackage{color}

\begin{document}

\title{Experimental evidence of low-lying gapped excitations in the quantum fluid at $\nu$=5/2}

\author{U. Wurstbauer}
	\email{uw2106@columbia.edu }
	\affiliation{Department of Physics, Columbia University, New York, NY 10027, USA}
\author{K. W. West}
	\affiliation{Electrical Engineering Department, Princeton University, Princeton, NJ 08544, USA}

\author{L. N. Pfeiffer}
	\affiliation{Electrical Engineering Department, Princeton University, Princeton, NJ 08544, USA}
\author{A. Pinczuk}
	\affiliation{Department of Physics, Columbia University, New York, NY 10027, USA}
       \affiliation{Department of Applied Physics and Applied Math, Columbia University, NY, New York 10027, USA}
\date{\today}

\begin{abstract}
The low-lying neutral excitation spectrum of the incompressible quantum Hall fluid at $\nu=5/2$ is investigated by inelastic light scattering. Gapped modes are observable only in a very narrow filling factor range centered at $5/2$ at energies that overlap estimates from transport activation gaps. The modes are interpreted as critical points in the wave-vector dispersion of excitations that preserve spin orientation. For very small changes  $|\delta\nu|\lesssim 0.01$ the gapped modes disappear and a continuum of low-lying excitations takes over indicating the transition from an incompressible fluid at 5/2 to a compressible state. Observations of spin wave modes indicate spin polarization of the 5/2 and 2+1/3 quantum Hall fluids.  
\end{abstract}


\maketitle

The impact of electron interactions in the first excited Landau level (N=1) differs significantly from those in the lowest (N=0) Landau level. While a compressible fermi sea is formed at filling factors  $\nu$=1/2 and $\nu$=3/2,  at $\nu$=5/2 and 7/2 incompressible fluids manifest by quantum Hall plateaus. The enigmatic states of the even denominator fractional quantum Hall effect (FQHE) states are the focus of intense research efforts. The Moore-Read (MR) Pfaffian many-body wave function is a prime candidate for the 5/2 state \cite{Moor1991}. In the MR description a weak attractive interaction among spin polarized fermion quasiparticles (QP) in the second LL  results in a condensed state of p-wave paired QPs. An exciting property of the MR state is that QP excitations obey non-Abelian braiding statistics \cite{Moor1991}, making the state a candidate to realize fault tolerant quantum computation \cite{Nayak2008,Stern2010}. Numerical evaluations consistently indicate that either the MR Pfaffian wave function \cite{Morf1998,Reza2000,Pete2008,Stor2011} or its particle-hole conjugate the Anti-Pfaffian (AP) wave function \cite{Lev2007,Lee2007,Reza2011} represent the $5/2$ ground state. Experimental observations support predictions of the MR or AP frameworks \cite{Radu2008,Dole2008,Will2009,Bid2010,Venk2011,Dole2011,Tiem2012,Stern2012}.  The state of spin polarization  of the 5/2 quantum fluid attracted much recent attention \cite{Tiem2012,Stern2012,Rhone2011b}, largely because full spin-polarization is a distinctive property of a non-Abelian state. 
\par
The low-lying excitation spectrum of the non-Abelian state is understood to support three types of neutral excitations: spin-wave modes (SW) of a fully spin-polarized FQHE ferromagnet \cite{Feig2009}, charge density  excitations that conserve spin (dispersive 'gap-excitation') and neutral fermions (NF's a.k.a. 'topological excitons') \cite{Bond2011,Moel2011,Sree2011}. 
Numerical calculations show that the wave vector dispersions of NF and gap excitations display distinct 'roton-minima' at finite  wave vector but converge to the same value in the large wave vector limit  \cite{Moel2011,Sree2011}.
\par
Neutral excitations that conserve spin (a.k.a. charge density excitations) exhibit characteristic wave-vector dispersions that are unique to each FQHE state. Inelastic light scattering experiments offer insights on the collective excitation spectrum of incompressible quantum fluids. Light scattering spectra  yield determinations of the $q \rightarrow 0$ mode \cite{Pinc1993} and of critical points in the wave vector dispersion that occur at roton minima and in the large wave vector limit ($q\rightarrow\infty$)  \cite{Cyrus2005,Rhone2011}. The charge density mode at $q\rightarrow\infty$ is regarded as the gap of the FQHE state \cite{Kallin}. There is consistent quantitative agreement between mode energies determined by inelastic light scattering experiments and calculated energies of neutral collective excitations of FQHE states in the  N=0 LL \cite{Rhone2011,Wurs2011}. Gap energies determined from activated transport are significantly smaller than calculated gap energies for realistic sample parameters \cite{Morf2003,Nueb2010,Samk2011,Pan2011}. The discrepancy is interpreted as arising from impact of residual-disorder on charge transport. 
\par
In this letter we report observations of low-lying excitations at $\nu$=5/2.  The neutral  modes are revealed in resonant inelastic light scattering (RILS) spectra. These are gapped  excitations in which the lowest modes have non-vanishing energy. The lowest energy band is the strongest in the RILS spectra. This band is interpreted as a roton critical point  in the $q$-dispersion of neutral charge density excitations. Remarkably, the gapped modes  are observed only in a very narrow filling factor range centered around 5/2. Very minor changes in filling factor of  $|\delta\nu |\lesssim 0.01$ result in a  transition from an incompressible quantum fluid with gapped excitations at 5/2 to compressible states with gapless excitations at filling factors only slightly away. These results provide experimental evidence that the intriguing incompressible quantum fluid that emerges at $\nu=5/2$ supports gapped low-lying excitations. 
\par
A mode that apears at the bare Zeeman energy $E_{Z}$ is assigned to a $q\rightarrow 0$ SW excitation. Its presence in the spectra reveals spin-polarization at $5/2$. However, the mode at $E_{Z}$ disappears for very small changes $\nu = 5/2 \pm \delta$ ($0.04 \geq \delta \geq 0.01$) indicating a rapid loss of spin polarization away from the incompressible state at 5/2.  Distinct SW modes at $E_{Z}$ are also observable in the range $2+1/3 \geq \nu > 2$, suggesting a high degree of spin polarization at those filling factors. These findings are in agreement with recent experiments showing a spin polarized ground state for 5/2 \cite{Tiem2012, Stern2012} and 2+1/3 \cite{Pan2012} and an unpolarized ground state at 2+2/3 \cite{Pan2012}.
\par
\begin{figure}[t]
\centering
\includegraphics[width=0.5\textwidth]{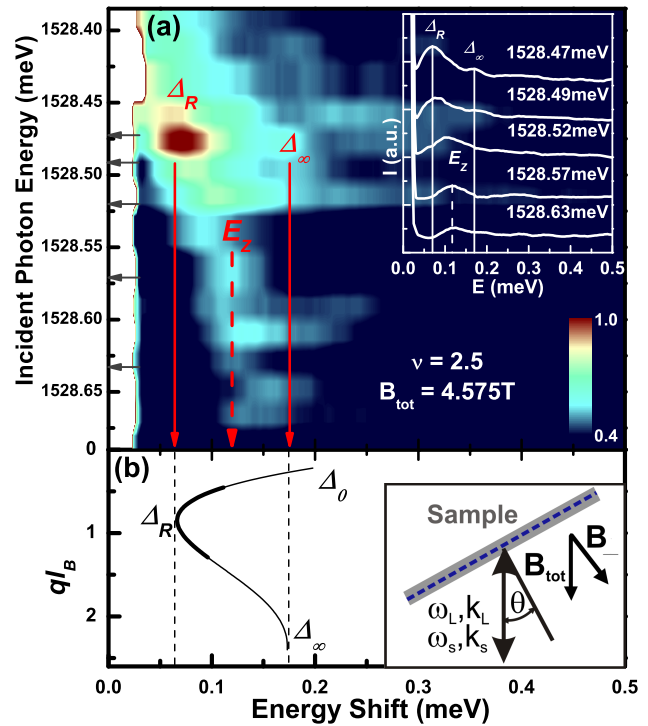}
\caption{(Color online) a) Color plot of RILS intensities in smoothed spectra measured at $\nu=5/2$ as function of $\omega_{L}$. Three modes are resonantly enhanced at the energies $\Delta_{R}$, $E_{Z}$ and $\Delta_{\infty}$. The spectra in the inset are at at $\omega_{L}$'s marked by horizontal arrows in the color plot.
b) Empirical wave vector dispersion based on the RILS mode energies showing at least one deep roton minimum in the neutral collective excitation spectrum.}
\label{fig:fig1}
\end{figure}

The ultra-clean two dimensional electron system is confined in a 300$\rm \AA$ symmetric single GaAs/AlGaAs quantum well.  Mobility and density after illumination are $\mu=23.9 \times 10^{6}\rm cm^{2}/\rm Vs$ and $n=2.9 \times 10^{11}\rm cm^{-2}$ at $T=300\rm mK$.  
The sample is mounted on the cold finger of a $^{3}$He/$^{4}$He-dilution dilution refrigerator with windows for direct optical access and inserted in the bore of a 16T superconducting magnet. All measurements were performed at $T\lesssim 45\rm mK$. The inelastic light scattering measurements are performed in a  back scattering geometry described in the inset to Fig. \ref{fig:fig1}  (b). There is a small tilt angle of $\theta=20$\textdegree~between the sample normal and the magnetic field. The impact of the small in-plane component of magnetic field on the 5/2 state is expected to be minor \cite{Xia2010}. The finite momentum transfer in back scattering is $k=|\vec{k_{L}}-\vec{k_{S}}|=(2\omega_{L}/c)\rm sin\theta$, where $k_{L(S)}$ is the in-plane component of the incident (scattered) photon , $\omega_{L}$ the incoming photon energy and $c$ the speed of light in vacuum. Spectra are excited by the linearly polarized light from a Ti:Sapphire laser.  $\omega_{L}$ is finely tuned to be close to the optical emission from the N=1 LL to resonantly enhance the light scattering intensity \cite{Rhone2011b}. The scattered light is dispersed by a triple grating spectrometer and recorded by a CCD camera with a combined spectral resolution of $<20\mu \rm eV$. The incident power density was kept well below $10^{-4} \rm W/\rm cm^{2}$.  
\par
\begin{figure}[t]
\centering
\includegraphics[width=0.5\textwidth]{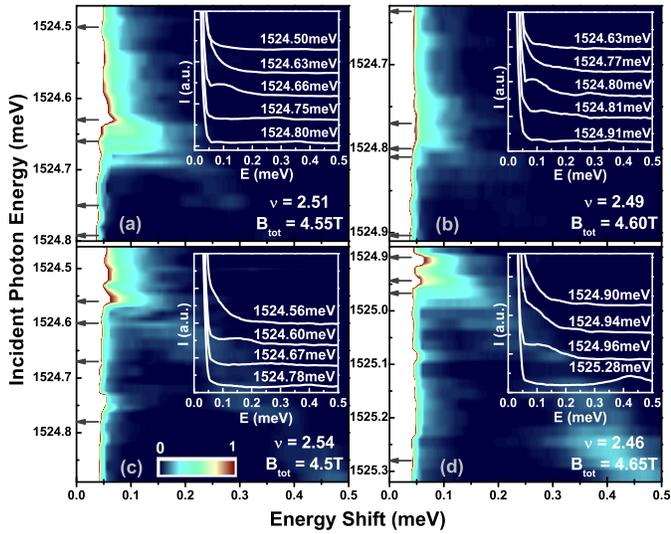}
\caption{(Color online) Color plot of RILS intensities measured at $\nu=5/2 \pm\delta$ ($\delta\leq 0.04$) as a function of $\omega_{L}$. The spectra in the insets are at at $\omega_{L}$'s marked by horizontal arrows in the color plot. These results illustrate the replacement of gapped excitations by gappless excitations slightly away from $\nu=5/2$}
\label{fig:fig2}
\end{figure}

Figure \ref{fig:fig1}  (a) displays RILS spectra at $\nu$=5/2 that show the evolution of the resonance enhacement of the scattering intensity.  There are three modes at positions that are highlighted by vertical lines in the color plot and also in the individual  spectra shown in the inset.  These are low-lying gapped excitations (i.e. at non-vanishing energy) of the 5/2 state. Figures \ref{fig:fig2}(a) and (b) show that by tuning the filling factor slightly away from 5/2 with a very small deviation of $|\delta\nu|\approx 0.01$ (change of the total magnetic field of $B_{tot}=25\rm mT$), the intensity of gapped low energy modes is greatly reduced, and there is evidence of a new, relatively weak, continuum of excitations that extends to vanishing energy. 
Such striking observations uncover a transition from an incompressible (gapped) quantum fluid at 5/2 to gapless compressible states at filling factors slightly away, and clearly identify the magnetic field of $\nu=5/2$.  
\par
Figures \ref{fig:fig2}(c) and 2(d) show that further away from $\nu=5/2$ ($|\delta \nu| \approx 0.04$) the RILS spectra consist of a strong  continuum of gapless excitations similar to RILS spectra reported by Rhone \textit{et al.} \cite{Rhone2011b}. In the range $2.9>\nu\geq 2.49$ the gapless excitations were attributed to a robust domain structure that emerges in that filling factor range \cite{Rhone2011b}. Our observation that a continuum of gapless modes is largely  suppressed at filling $\nu$=5/2 suggests that the phase of the incompressible quantum fluid at 5/2 overwhelms the non-uniform compressible phases that are predominant at filling factors slighly away from 5/2.
\par
The strongest band in Fig.\ref{fig:fig1} (a) is the lowest mode labeled $\Delta_{R}$ that is centered at $0.07\rm meV$. There are two weaker features in the spectra. One, at energy  $0.12\rm meV$ is at the bare Zeeman energy $E_{Z}$ of GaAs. The other one at $\Delta_{\infty}=0.17\rm meV$ is the highest energy mode in these spectra. The band of the mode at $\Delta_{R}$ overlaps the range of gap energies reported in activated magneto transport at $\nu=5/2$ \cite{Choi2008,Dean2008,Nueb2010,Kumar2010,Samk2011,Pan2011}. The similarity between lowest mode energies seen in RILS and activation energies strongly suggests that the modes at $\Delta_{R}$ are low-lying spin-conserving neutral excitations of the incompressible quantum fluid at $\nu=5/2$.   
\par
The excitations in Fig.\ref{fig:fig1} (a)  are interpreted with the conceptual framework that was successful for RILS by quantum Hall fluids of the N=0 LL. For FQHE states with $\nu<1$ RILS spectra display wave vector conserving modes at $k=q\rightarrow 0$ \cite{Pinc1993}. There are stronger modes at large $q$ ($q>k$) that are activated by breakdown of wave vector conservation due to weak residual disorder. These RILS spectra display maxima at critical points in the mode dispersion \cite{Cyrus2005,Rhone2011,Wurs2011}. 
 \par
To extend this framework to the quantum Hall state at $\nu=5/2$ we show in Fig.\ref{fig:fig1} (b) a tentative schematic wave vector dispersion of low-lying gapped excitations. This dispersion displays a characteristic magnetoroton minimum at wave vector $q_{R}\approx 1/l_{B}$, where $l_{B}=[eB/(\hbar c)]^{1/2}$ is the magnetic length \cite{Bond2011,Moel2011,Sree2011}. Within this framework the lowest mode $\Delta_{R}$ is interpreted as the critical point that occurs at the roton. The highest energy mode $\Delta_{\infty}$ is identified as the large $q$ limit of the mode dispersion. This limit represents a neutral particle-hole pair at large separation that is understood as the intrinsic transport gap of the $\nu=5/2$ state. The energy $\Delta_{\infty}=0.17 \rm meV$ (see Fig. \ref{fig:fig1} ) is in very good agreement with a calculation of the intrinsic gap that includes finite width and LL mixing \cite{Morf2003}. The value of $\Delta_{\infty}$ also agrees well with  experimental determinations of the  intrinsic transport gap (from activated transport) that take into account the impact of residual-disorder by a subtractive broadening term $\Gamma$ \cite{Samk2011}. 
\par
\begin{figure}[t]
\centering
\includegraphics[width=0.5\textwidth]{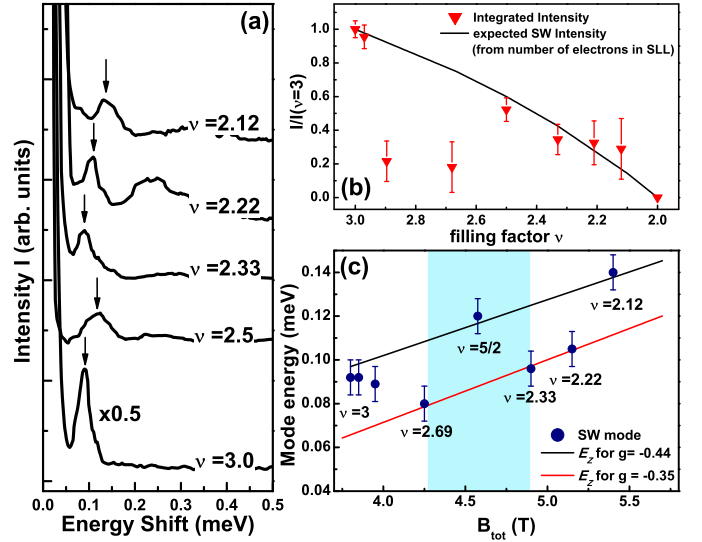}
\caption{(Color online) a) Spin wave modes at $E_{Z}$ for states in the range 2.5 $\geq \nu \geq$ 2 and for $\nu = $ 3. b) Integrated SW mode intensity after background substraction relative to the intensity of the SW at $\nu =3$. The solid line depicts the population of electrons residing in the lower spin-branch of the SLL. The solid triangles are the measured integrated mode intensities. c) SW mode energy as a function of total magnetic field displaying an effective $g$-factor of about $|g|\approx 0.44$ at 5/2 and around integer filling factors and $|g|\approx 0.35$ close to odd-denominator filling factors $\nu$=2+2/3, 2+1/3, 2+1/5.}
\label{fig:fig3}
\end{figure}

 The energy of the band at $E_{Z}$ in Fig.\ref{fig:fig1} (a) is written as $E_{Z}=\mu_{B}gB_{tot}$, where $\mu_{B}$ is the Bohr magneton and $g$ is a bare g-factor. $E_{Z}=0.12\rm meV$ yields $|g|$=0.44, close to the bare $g$-factor of GaAs. This spin excitation can be regarded as the $q\rightarrow 0$ limit of a dispersive SW mode as required by Larmor's theorem. To explore this interpretation, which implies that the quantum fluid at $5/2$ has spin polarization, we show in Fig. \ref{fig:fig3}(a) spectra where a band at $E_{Z}$ is present for $\nu\leq 5/2$. The SW spectrum in the fully spin-polarized state at $\nu=3$ is also shown. For filling factors $\nu<3$ the lowest spin-up branch of the N=1 LL is depopulated and the expected intensity of the SW quickly drops. Figure \ref{fig:fig3}(b) shows measurements of the integrated intensity at $E_{Z}$ relative to the integrated intensity of the $E_{Z}$ band at $\nu=3$. There is good agreement of the measured points at 5/2 and $2+1/3\leq \nu < 2$ with a calculation of the drop in population of the lowest spin-up branch of the SLL shown in Fig.\ref{fig:fig1} (b) as a solid line. The finding that the intensity of the SW at $E_{Z}$ is linked to the expected population of the spin-up branch of the of the N=1 LL suggests that FQHE states at $\nu= 5/2$ and $2+1/3$ could have full polarization of spin.  

\par
The absence of SW modes in the near vicinity of 5/2 and the emergence of a strong scattering continuum of gapless low-lying excitations for $\nu=3 \pm \delta$ resemble the behavior of the SW mode at and around the quantum Hall ferromagnet $\nu=1$ \cite{Gallais2008}. The findings at $\nu=1 \pm \delta$ were interpreted as evidence of Skyrme textures in the ground state \cite{Gallais2008}. While the formation of Skyrmions at 5/2 has been numerically investigated \cite{Wojs2010}, changes such as the absence/reduction of the SW intensities could also be caused by the fluid at 5/2 breaking-up into domains thus suppressing the intensity of collective excitations. In Fig. \ref{fig:fig3}(a) the SW modes measured at $\nu=5/2$ and at $\nu=2.12$ are significantly broader  than those at the other filling factors. The enhanced broadening could be a signature of a more complex interplay of interactions and onset of localization from non-uniform domains. Other plausible mechanisms could emanate from residual QP interactions \cite{Wurs2011}.
\par
In Fig. \ref{fig:fig3}(c) we plot SW mode energies as a function of $B_{tot}$ in the filling factor range $3 \geq \nu \geq 2$. The observed energies fall on two lines with slightly different values of $|g|$-factor suitable for GaAs. At filling factors 5/2, 3 and 2.12, $|g|\approx 0.44$. For odd denominator filling factors $\nu=2+2/3, 2+1/3, 2+1/5$ $|g|\approx 0.35$. There is a clear trend towards a reduction in g-factor as $\nu<3$, but the recovery of $g$-factor value at 5/2 is intriguing. 
\par
In summary, we report the discovery of gapped low-lying excitations in a narrow range of filling factor around $\nu=5/2$. The resonant inelastic light scattering experiments reveal that for filling factor changes $|\delta\nu |\lesssim 0.01$ the gapped excitations are replaced  by gapless modes. The experiments also show spectra in which the observation of a SW mode at the bare Zeeman energy is consistent with full spin-polarization at 5/2 and also in the filling factor range $2+1/3 \geq \nu > 2$. These findings  demonstrate new experimental venues to study intriguing FQHE fluids that could have applications in topological quantum information processing.

\textbf{Acknowledgments} We would like to thank J.K. Jain, G. J. Sreejit and A. W\'{o}js for insightful discussions. The work at Columbia is supported by the National Science Foundation (NSF) (DMR-08034445 and CHE-0641523).  The work at Princeton was partially funded by the Gordon and Betty Moore Foundation as well as the NSF MRSEC Program through the Princeton Center for Complex Materials (DMR-0819860). U.W. acknowledges partial support from the Alexander von Humboldt Foundation.

\nocite{*}

\end{document}